\documentclass{aa}
\usepackage{siunitx}
\usepackage{bbm}
\usepackage{natbib,twoopt}
\bibpunct{(}{)}{;}{a}{}{,} 
\usepackage[breaklinks=true]{hyperref} 
\makeatletter

\newcommandtwoopt{\citeads}[3][][]{\href{http://adsabs.harvard.edu/abs/#3}%
{\def\hyper@linkstart##1##2{}%
\let\hyper@linkend\@empty\citealp[#1][#2]{#3}}}
\newcommandtwoopt{\citepads}[3][][]{\href{http://adsabs.harvard.edu/abs/#3}%
{\def\hyper@linkstart##1##2{}%
\let\hyper@linkend\@empty\citep[#1][#2]{#3}}}
\newcommandtwoopt{\citetads}[3][][]{\href{http://adsabs.harvard.edu/abs/#3}%
{\def\hyper@linkstart##1##2{}%
\let\hyper@linkend\@empty\citet[#1][#2]{#3}}}
\newcommandtwoopt{\citeyearads}[3][][]%
{\href{http://adsabs.harvard.edu/abs/#3}
{\def\hyper@linkstart##1##2{}%
\let\hyper@linkend\@empty\citeyear[#1][#2]{#3}}}
\makeatother

\newcommand{\code}[1]{\texttt{#1}}

\begin{document}
\title{ASTROMER}

\subtitle{A transformer-based embedding for the representation of light curves}

\author{C. Donoso-Oliva\inst{1,3,4}
     \and I. Becker\inst{2,4,5}
     \and P. Protopapas\inst{3,5}
     \and G. Cabrera-Vives\inst{1,4}
     \and Vishnu M.\inst{5}
     \and Harsh Vardhan\inst{5}}

\offprints{Cristobal Donoso-Oliva, \email{cridonoso@inf.udec.cl}}

\institute{
Department of Computer Science, Universidad de Concepcion, Concepcion, 4070386, Chile
  \and 
  Department of Computer Science, Pontificia Universidad Catolica de Chile, Macul, Santiago 7820436, Chile
   \and
Inst. for Applied Computational Science, Harvard University, Cambridge, MA 02138, USA
  \and 
Millennium Institute of Astrophysics (MAS), Nuncio Monsenor Sotero Sanz 100, Providencia, Santiago, Chile
\and
Univ.AI, Singapore, 050531, Singapore
}

\date{Received x xxx xxxx / Accepted x xxx xxxx}

\abstract{
Taking inspiration from natural language embeddings, we present ASTROMER, a transformer-based model to create representations of light curves. ASTROMER was pre-trained in a self-supervised manner, requiring no human-labeled data. We used millions of R-band light sequences to adjust the ASTROMER weights. The learned representation can be easily adapted to other surveys by re-training ASTROMER on new sources. The power of ASTROMER consists of using the representation to extract light curve embeddings that can enhance the training of other models, such as classifiers or regressors. As an example, we used ASTROMER embeddings to train two neural-based classifiers that use labeled variable stars from MACHO, OGLE-III, and ATLAS. In all experiments, ASTROMER-based classifiers outperformed a baseline recurrent neural network trained on light curves directly when limited labeled data was available. Furthermore, using ASTROMER embeddings decreases computational resources needed while achieving state-of-the-art results. Finally, we provide a Python library that includes all the functionalities employed in this work. The library, main code, and pre-trained weights are available at \href{https://github.com/astromer-science}{https://github.com/astromer-science}
}

\maketitle
\section{Introduction} \label{sec:intro}
Over the past decades, efforts have been made to develop machine learning tools to analyze and discover variable phenomena in the sky. These tools will face a challenge with the construction of the new generation of telescopes, which will generate substantially more data than the old generation \citepads{kremer2017big}. Moreover, the observations will be deeper and more precise than ever.

With the upcoming telescopes such as the Vera C. Rubin Observatory \citepads{2019ApJ...873..111I}, periodic observations of the a significant portion of the sky every few days will become the norm. They will produce light curves for all the objects in the observed fields.

Traditional machine learning methods rely on features to explore the variable behavior of light curves. They are based on quantities such as amplitude, period, and color information, among others \citepads{sanchez2021alert}. The expected volume of data will present a significant challenge to feature-based methods if applied to every measured object.

Deep learning techniques have an advantage over traditional machine learning as they do not need to pre-calculate features, as they extract informative representations of the data automatically \citepads{lecun2015deep}. Also, deep learning methods leverage GPU parallelization, which accelerates information extraction from the light curves.

In the last years, some of these methods have been developed with promising results. \citetads{naul2018recurrent} presented a recurrent autoencoder to extract features from folded periodical light curves. After training, the authors fitted a Random Forest classifier on the embedded space, outperforming models trained on pre-defined features. This model was one of the first approaches in astronomy that used unlabeled light curves to extract characteristics from an unsupervised deep learning model.
A similar idea was proposed a year later, by \citetads{tsang2019deep} which used an autoencoder not only for classification but also for novelty detection. 
These approaches remain supervised, which is problematic for training on small datasets (\citetads{charnock2017deep}).

Self-supervised models are ideal to overcome the limitations of labeled datasets, as they can generate the target values automatically without human-generated labels (\citetads{liu2021self}). Most of these methods learn representations by solving auxiliary tasks which do not require any label, such as infilling of time series. The learned representations can be used downstream to solve other tasks, such as classification of variable stars or regression of physical parameters such as temperature or redshift. This learning scheme has achieved impressive results in Natural language processing (NLP), with Bidirectional Encoder Representations from Transformers (BERT, \citetads{devlin2018bert}) being one of the most notable examples. 
BERT-like models learn to extract contextual representations using two auxiliary tasks from a dataset of raw text. In this stage, known as pre-training, the model learns a general representation of the data, without being specific to any particular task. This process is the most resource-intensive part that usually takes weeks to complete, as the models as well as the datasets are large. 
After this stage is complete, the model only needs a few hours of further training to adjust the weights to a more specific domain, in a stage known as fine-tuning. 
Once finalized, BERT is used as the initial stage of a new model. This scheme enables BERT-based models to achieve state-of-art performance. 

The main components of BERT are the self-attention layers (\citetads{vaswani2017attention}), that codify similarities between words in the input sentence. These layers can be computed efficiently in parallel, in contrast to the sequential behavior of recurrent neural networks. Recently, some attention-based works have been proposed in astronomy, showing competitive results with the state-of-the-art (\citeads{allam2021paying}, \citeads{pimentel2022deep}, \citeads{pan2022astroconformer}).

Following the advances on NLP to treat sequential data, representation learning and self-supervised training strategies, we present ASTROMER, a self-supervised model to extract a general representation of astronomical light curves.

Our contributions can be summarized as follows:
\begin{enumerate}
    \item We introduce ASTROMER, a self-supervised model that creates light curves representations taking advantage of the massive unlabeled volume of astronomical data.
    \item Using ASTROMER representations can match or decrease the number of epochs needed to train downstream models, such as classifiers of variable stars.
    \item We empirically demonstrate the benefits of using pre-trained representations when training classifiers on small datasets. Using less than 100 samples per class, we significantly overcome an Long-short term memory (LSTM) trained on light curves \textbf{directly}.
    \item We provide a python library that includes pre-trained models and the necessary infrastructures to fine-tune and obtain domain specific embeddings.
    \item Sharing pre-trained models by the community to save computational resources, decreasing C02 emissions while improving the performance of automatic learning models.
\end{enumerate}
Furthermore, we aim to create different versions of ASTROMER, just as BERT has many variations depending on the target (\citetads{polignano2019alberto},
\citetads{liu2019roberta}, \citetads{de2019bertje}, \citetads{moradshahi2019hubert}, \citetads{masala2020robert}, \citetads{vunikili2020clinical}).

\section{Problem Statement}\label{sec:astromer}
Let $\rm{X} \in \mathbb{R}^{L \times d_x}$ be a single-band light curve where $d_x$ is the number of features with $L$ observations over time. In this case, every observation consists of $d_x=2$: the magnitude and the Modified Julian Date (MJD), where the observations occur. Naturally, the number of observations and times vary between different stars, and it strongly depends on the survey science goals. In this representation, the maximum number of points in the light curves remains fixed even if some of them are shorter than $L$, in which case we perform zero padding and masking. 

The main objective, is to train a model $f(X, \mathcal{D}_A, \theta)$ on a massive set of light curves $\mathcal{D}_A$ from a given survey $A$, with trainable parameters $\theta$. In particular, we propose to use learned representations of a transformer-based encoder to create embeddings $\rm{Z} \in \mathbb{R}^{L \times d_k}$ representing the objects' variability in $d_k$-dimensional space. We can fine-tune the model's weights to adapt to other surveys and use it to solve downstream tasks, such as classification or regression. 

\section{Methods}
In this section, we describe the main components of ASTROMER, which belong to the Transformer neural network proposed by \citetads{vaswani2017attention}. In particular, we focus on two processes: the self-attention block and the positional encoding (PE). Both methods are effective to capture relationships between observations to encode light curves.

\subsection{Self-Attention Block}\label{subsection:attention}
\citetads{vaswani2017attention} introduced the self-attention mechanism as an alternative to the classical attention technique based on Recurrent Neural Networks (RNNs). The idea is to quantify relationships between observations without conditioning the operations to follow a sequential order. Thus, unlike RNNs, self-attention blocks can be executed in parallel, being more efficient and faster to train.

An attention block consist of multiple self-attention heads, that compute the similarities of every input within the sequence to each other. In other words, every head measures the relationship between pairs of observations; the more an observation affects the other, the more attention the model pay. As shown in Equation \ref{eq:attention}, we perform a weighted sum over the input values $\rm{V}_\textit{i}$ to obtain the attention matrix $\rm{Z}_\textit{i}$ using learnable query-key compatibilities ($\rm{Q}_\textit{i}\rm{K}_\textit{i}^{\top}$).
\begin{equation}\label{eq:attention}
    \rm{Z}_\textit{i} = \textnormal{softmax}\left(
    \frac{\rm{Q}_\textit{i}\rm{K}_\textit{i}^{\top}}
         {\sqrt{d_k}}\right) \rm{V}_\textit{i},
\end{equation}
Queries, keys, and values ($\rm{Q}_{\textit{i}}, \rm{K}_{\textit{i}}, \rm{V}_{\textit{i}}$) are input transformations such that,
\begin{eqnarray}\label{eq:querykeyvalue}
    \rm{Q}_i = \rm{X}{\rm W}_\textit{i}^q,\hspace{2mm}
    \rm{K}_i = \rm{X}{\rm W}_\textit{i}^k,\hspace{2mm}
    \textnormal{and} \hspace{2mm}
    \rm{V}_i = \rm{X}{\rm W}_\textit{i}^v.
\end{eqnarray}
with $\rm{W}_\textit{i}^q, \rm{W}_\textit{i}^k$ and $\rm{W}_\textit{i}^v$ trainable weight matrices of the $i$-th head and $d_k$ a hyper-parameter specifying the embedding size of each self-attention head. The normalization factor $\sqrt{d_k}$ in the denominator scales the variance of the product, while the softmax bounds the values to represent a multinomial probability distribution. 

\begin{figure}
    \centering
    \includegraphics[scale=0.88]{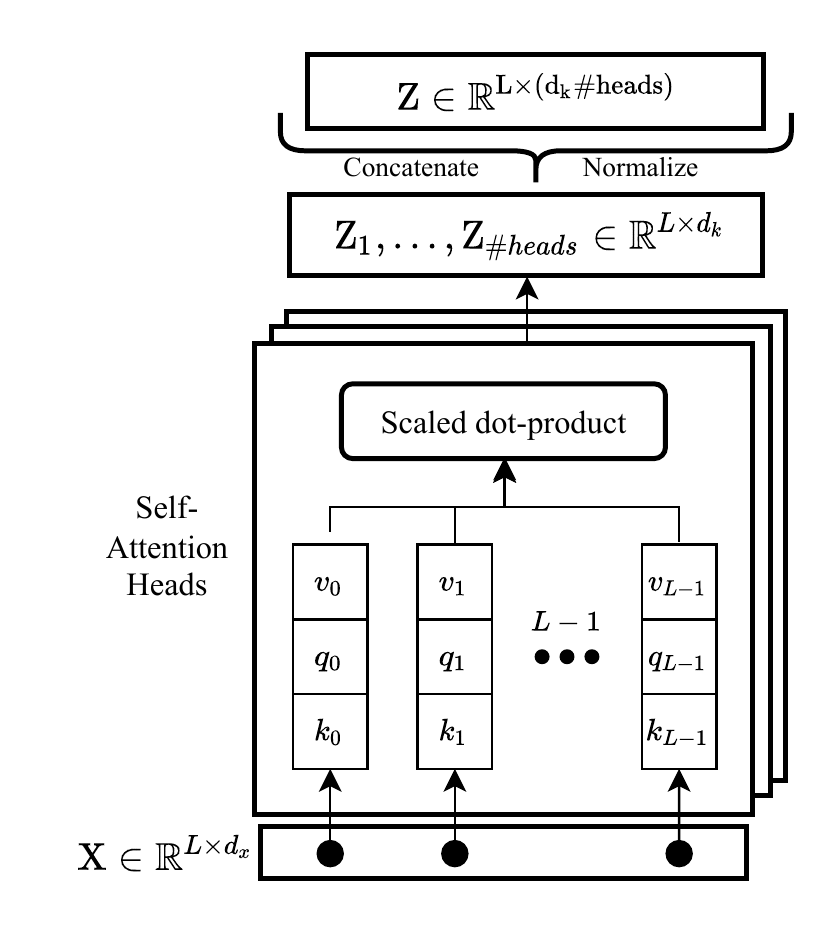}
    \caption{Self-attention block diagram. Each observation vector $\textbf{x}_l \in \rm{X}$ (denoted by solid circles) is projected into three vectors: (k)ey, (q)uery, and (v)alue. Then their similarities are computed by the scaled dot-product between vectors according to Equation \ref{eq:attention}.}
    \label{fig:self_att_block}
\end{figure}

The final attention vector is the normalized concatenation of $\{\rm{Z}_0,..,Z_i,...,Z_{\text{\#heads}}\}$ the output from every $i$-th head in the block. Figure \ref{fig:self_att_block} illustrates the self-attention block.

\subsection{Positional Encoding (PE)}\label{sec:pe}
Light curves are sequences of brightness measurements as a function of time, generally not equally spaced. The cadence of a survey is given by the time interval between observations, which is often non-periodic.

Temporal information is essential to characterize light curves since they bring information about the variability of a particular object. For example, RR Lyrae and Delta Scuti stars have short periods, ranging from hours to days, while the periods of type II Cepheids and Long Period Variables can be months or years.

By construction, self-attention layers do not take into account the temporal information of light curves when learning similarities between the observations. In other words, any re-ordering of the magnitudes within a sequence would produce the same attention vector. 

A traditional way to include temporal information during self-attention learning consists of modifying the input to incorporate the relative positions into the same vector. We create a representation for the time and add that information to the brightness representation. Note that the positions range from 0 to L-1, where L is the length of the light curve.

In this work, we adjust a classical positional encoder used by \citeads{vaswani2017attention}, to work with the time domain of the light curves\footnote{We based our implementation on tensor2tensor code (\citeads{tfpe}) as it achieved a better validation RMSE after pre-training than using the original implementation}. The PE projects an embedding that encodes relative positions and it is consistent to any transformation of them, no matter the number of observations or the first day in MJD where the object began to be observed. It consists of trigonometric functions that codify each observational time $t_l$ directly, from the $l$-th observation at different $w_j$ frequencies.

\begin{equation}\label{eq:pe}
PE_{\textit{j}, t_{l}} = \left\{
        \begin{array}{ll}
            \sin{(t_{l}\cdot\omega_{\textit{j}})} & \quad j \text{ is even} \\
            \cos{(t_{l}\cdot\omega_{\textit{j}})} & \quad j \text{ is odd}
        \end{array}
    \right.
\end{equation}

In Equation \ref{eq:pe}, $\textit{j} \in [0,...,d_{pe}-1] $, where $d_{pe}$ is the PE dimensionality and $w_\textit{j}$ the angular frequency defined as
\begin{equation}\label{eq:angular_freq}
    \omega_{\textit{j}} = \frac{1}{1000^{2\textit{j}/d_{pe}}}.
\end{equation}

\begin{figure}
    \centering
    \includegraphics[scale=0.85]{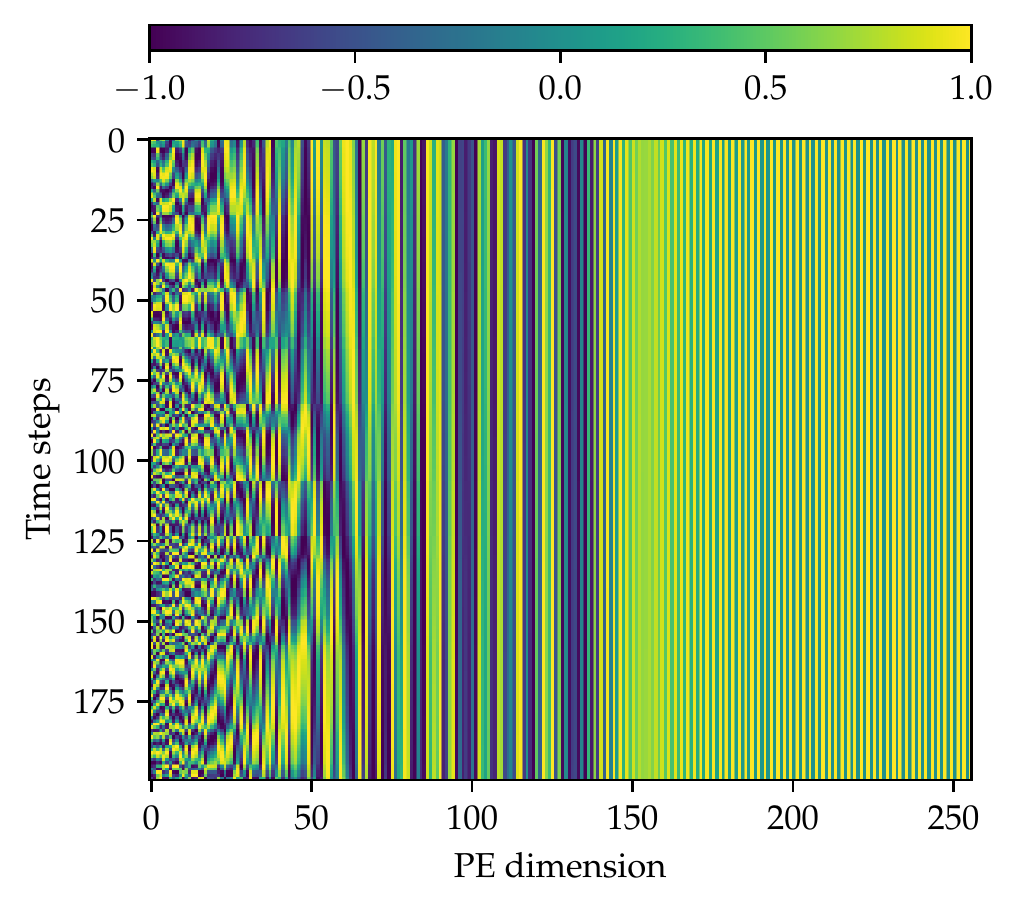}
    \caption{Resulting positional embedded space from the F\_1.4175.3433 MACHO object.}
    \label{fig:pe}
\end{figure}

Trigonometrical functions are a natural choice to capture periodic behavior while bringing unique values within $[-1, 1]$, which work efficiently on GPUs (\citetads{patel2020designing}). The sine and cosine functions take $d/2$ different angular frequencies, spanning wavelengths from $2\pi$ to $2\pi\times1000^2$. We make the same remarks as in the original paper regarding the function used. A non-trainable PE is enough to achieve reasonable results, simplifying the network. The value of $1000$ gave the best results, rather than the original $10000$.
The angular frequencies are indexed by the PE dimension number. Larger frequencies are given by the smaller PE dimensions while the smaller frequencies are given by the larger PE dimensions. 
Figure \ref{fig:pe} shows the resulting PE given the sequence of times of the MACHO light curve F\_1.4175.3433. At high frequencies, the PE shows high variability across all the time steps, which shows dynamic behavior. In contrast, lower frequencies show low variability.

The PE relates directly to the objects under study. Since the periods of variable stars typically range from 0.1 to 1000 days \citepads{catelan2015pulsating} the PE should be sensible to periods in this range. To capture short periods, the minimum wavelength should be $2\times 0.1$. In practice, the survey's cadence conditions the shortest period we can study.
Moreover, most objects will not show variability for periods larger than 1000 days. As shown in Figure \ref{fig:pe}, the PE doesn't show any change across the time-steps.

\section{Proposed Solution}
Here we introduce and describe ASTROMER, a transformed-based model capable of producing embeddings from light curves. Embeddings summarize photometric data according to a learned representation that can be easily adapted to new data by retraining (or fine-tuning) the model's weights. In particular, we take inspiration from BERT (\citetads{devlin2018bert}), a self-supervised trained model that does not need human-based annotations to learn representations of the input data.

In the following sections, we explain the main modules and objectives of the proposed model. Before describing the component of the model, we motivate the architecture by describing the target task in Section \ref{sec:MLM}. Next, Section \ref{sec:architecture} presents a general overview of the architecture, where we introduce the ASTROMER modules. Finally, the last sections detail the inner mechanisms of encoding (Section \ref{sec:building_rep}) and decoding (Section \ref{sec:defining_decoder}) light curves.

\subsection{Target task: Modeling of masked light curves}\label{sec:MLM}
ASTROMER aims to learn representations that summarize and characterize the light curve's domain. It does by reconstructing the input light curve from a sub-sample of observations. The embedding should be as general as possible to help downstream tasks, such as classification or regression. In this sense, the model needs to adjust its weights to create a vector that is informative enough to recover the input light curve.

In order to force the model to learn contextual information, we hide a random subset of magnitudes within the sequence, and reconstruct them using the remaining part of the light curve. Hidden elements are informed to the model by a mask vector $\boldsymbol{m_i}$ that is different for every $i$-th light curve sample. The loss function then takes the form,
\begin{eqnarray}\label{eq:loss}
    \mathcal{L}oss = \sqrt{\frac{1}{N-1}\sum_{i=0}^{N-1}\sum_{l=0}^{L-1} m_{il}(x_{il} - \hat{x}_{il})^2},
\end{eqnarray}
where $N$ is the number of examples, and $L$ is the length of the input sequences, equal for all examples.
Notice that the loss function containing the actual observation $x_{il}$ and the prediction $\hat{x}_{il}$ is equivalent to the root-mean-square error (RMSE). In equation \ref{eq:loss}, the mask vector $\boldsymbol{m}_{i}$ only allows summing over the masked values' error; nevertheless, the model reconstructs the entire sequence.

\subsection{Model Architecture}\label{sec:architecture}
\begin{figure}
    \centering
    \includegraphics[scale=0.9]{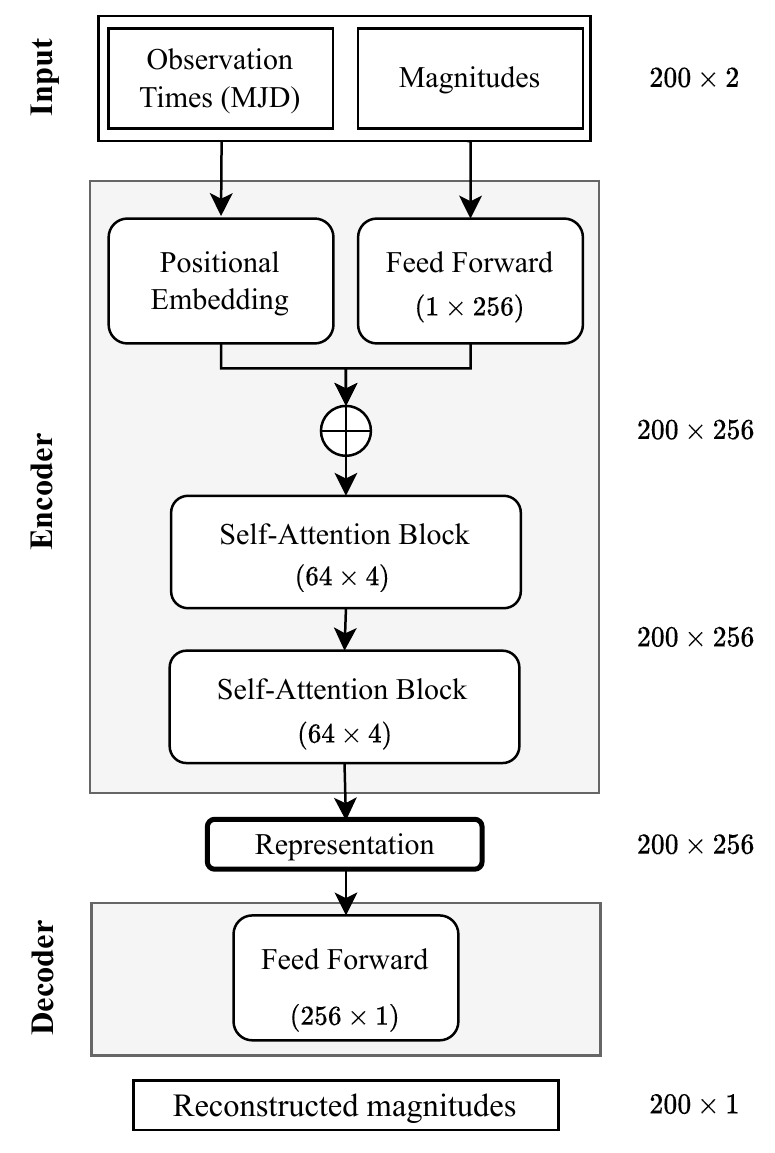}
    \caption{ASTROMER architecture}
    \label{fig:arch}
\end{figure}

ASTROMER has an encoder-decoder architecture, as shown in Figure \ref{fig:arch}. Input samples are fixed-length light curves of $L=200$ observations with two descriptors each: observation times and magnitudes.
The first part of the encoder is tasked to compute the PE, which receives the time values at each step and projects them into vectors of 256 dimensions.  
The next step consists of adding magnitudes to the PE without interfering with the temporal encoding. As shown in Figure \ref{fig:arch}, we train a feed forward network (FNN) without hidden layers that transform each magnitude to a vector of size 256. Notice the dimension of the FFN match the size of the positional embedding. In order to not alter the temporal encoding, the FNN should learn to project brightness information in a way that can be summed with the constant part of the PE, i.e., after the 100-th dimension in Figure \ref{fig:pe} \footnote{We confirm the assumption in Appendix \ref{ap:input_pe}, where the model -after training- learned to project magnitudes on the $\sim 200\text{-}$th dimension.}. After adding the projected magnitudes and the PE, the new input is a matrix of $200 \times 256$.

The core of ASTROMER takes place on the two self-attention blocks. Each block contains four heads with 64 neurons each. The final representation is the normalized concatenation of the heads of the last block. Then, the resulting embedding is a collection of 200 vectors of size 256, describing the attention of every observation to each other. Notice that $d_{pe} = d_{k}\#heads = 256$.

In the decoder, we take the representation to build the model's output, which during the pre-training consist of a linear feed-forward network with no hidden layers that reconstructs the input magnitudes. Although the decoder presented in Figure \ref{fig:arch} is exclusive for pre-training ASTROMER, we can use different decoding layers focusing on other downstream tasks.
%
%
\subsection{Learning representations}\label{sec:building_rep}

ASTROMER uses self-attention blocks to transform light curves into embeddings via learnable parameters. As mentioned in Section \ref{sec:MLM}, we train ASTROMER to predict a subset of masked observations using the concatenated attention vector at the end of the encoder. The embeddings must capture enough information to reconstruct magnitudes using only the surrounding local context of the hidden observations.

In order to exclude the masked observations from the self-attention blocks, we make zeros on all the dimensions associated with the hidden elements. Therefore, by modifying Equation \ref{eq:attention} we obtain,
\begin{equation}\label{eq:masked_attention}
    \rm{Z}_\textit{i} = \textnormal{softmax}\left(
    \frac{\rm{Q}_\textit{i}\rm{K}_\textit{i}^{\top} + (-10^9)\rm{M}}
         {\sqrt{d_k}}\right) \rm{V}_\textit{i},
\end{equation}
where $\rm{M}$ is a binary matrix that is 1 on masked positions and 0 otherwise. The $\rm{M}$ matrix is dynamically created before the encoder. Notice that for the points in the light curve that are hidden (i.e. for which \rm{M} is 1), the value of the softmax function will be close to zero, making $\rm{Z}_i$ close to zero. In other words, when the observation is masked, its assigned predicted value will be zero, which represents no attention.

\begin{figure}[!h]
    \centering
    \includegraphics[scale=0.75]{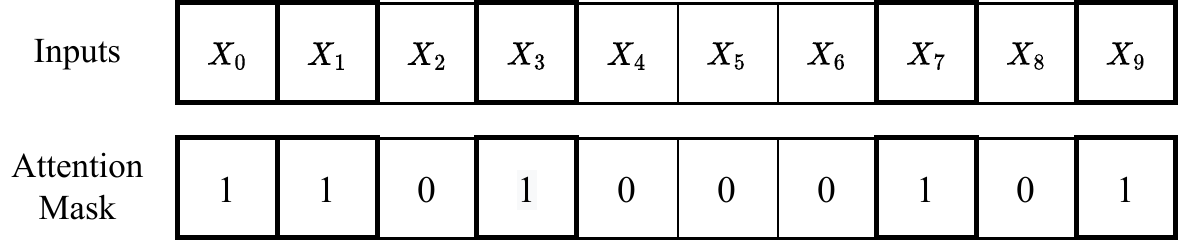}
    \caption{
Preliminary input example with 50\% of masking. ASTROMER aims to predict masked observations (denoted by 1's) using only the attention associated with zero-tagged elements.
    }
    \label{fig:input_m_0}
\end{figure}

In this work, we mask 50\% of the total number of observations per light curve. Notice $\rm{M}$ should be reshaped to match the square matrix $\rm{Q}_\textit{i}\rm{K}_\textit{i}^{\top} \in \mathbb{R}^{L\times L}$.  Figure \ref{fig:input_m_0} shows an example of a 10-observations input.

\begin{figure}[!h]
    \centering
    \includegraphics[scale=0.75]{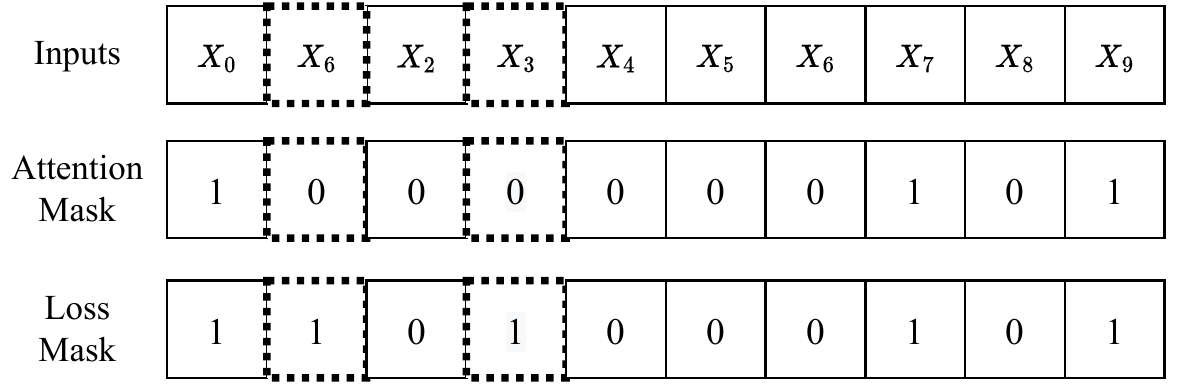}
    \caption{
Final input composition. Following the example in Figure \ref{fig:input_m_0}, 20\% of the masked values are replaced by random magnitudes while changing the 1's in the attention mask vector to 0's. Similarly, we make another 20\% of the masking visible, keeping the actual observations. Doted line squares indicate both random and real observations. At this point, we should keep a second mask containing the initial 50\% masked values to evaluate the loss function.
    }
    \label{fig:input_m_1}
\end{figure}

During training, the mask $\rm{M}$ provides the model with information about the observations to be predicted. The model can only focus on the target magnitudes and forget contextual information about unmasked inputs. To avoid learning biased representations conditioned by the masking information only, \citeads{devlin2018bert} proposed to replace a subset of masked values with random and real elements. In practice, we turn 20\% of the masked values from 1 to 0 while replacing its associated magnitudes with random values from the same sequence. Similarly, we make another 20\% of the masked values visible but this time without changing their magnitudes. The unmasked values are still considered in the loss function as they are part of 50\% of the initial masked elements. Figure \ref{fig:input_m_1} shows the final composition of the input.

Having defined the mask vector, we initiate the forward pass of the encoder by transforming input times and magnitudes to a 256 vector that mixes temporal and brightness information. Finally, we pass the combined input sequences via two self-attention blocks that capture similarities between observations and then output the attention-based embeddings.

%
%
\subsection{Decoding embeddings}\label{sec:defining_decoder}
The decoder is essential to adjust the encoder weights. The output conditions the purpose of the embedded representation. Despite considering only masked values in the loss function, the decoder reconstructs the entire sequence of magnitudes. As we explained in Section \ref{sec:MLM}, using an mask vector, we give a value of zero to the observations that are not part of the loss.

The embedding is a collection of 200 vectors (the length of the input sequence) of size 256 (model dimensionality). Every vector describes the relationship between a particular observation and the rest of the sequence. In order to recover the original magnitude, we apply a feed-forward network without hidden layers and no activation on all the vectors of size 256 separately. Although we transform every vector separately, the network weights remain the same.

Finally, it is important to mention that the decoder only works during the pre-training and fine-tuning of ASTROMER. After that, we usually use the encoder along with a new decoder specializing in another specific task, such as the classification of variable stars. At this step, the encoder can also be trained together with the new decoder. However, the encoder cannot change drastically, as it would forget the learned representations.
\section{Data description}\label{sec:data}
This section describes the data involved in the pre-training and fine-tuning of the ASTROMER model. 
Every light curve contains magnitudes and observation times in Modified Julian Date (MJD).

ASTROMER takes advantage of a massive set of light curves, not necessarily labeled. However, to evaluate a downstream task (i.e., classification), we use catalogs of variable stars with their corresponding labels. This work simulates the science case of having a small subset of labeled samples and employs them to fine-tune the model even when the class information is unnecessary.
%
%
\subsection{Unlabeled dataset}\label{sec:unlabeled}
R-band light curves of the Massive Compact Halo Object survey (MACHO; \citetads{alcock2000macho}) were collected from the Galactic Bulge (Fields 1 and 10), and the Large Magellanic Cloud (LMC, Fields 101-104). The photometry was taken as it is provided, without doing any additional processing.

Given the nature of photometric observations, the observed flux of every star will exhibit some kind of variability, such as the intrinsic observational noise. As such, we cannot expect every star to be variable. 
Even though non-variable and noisy data can be helpful to regularize weights (\citetads{bishop1995training}), we discard some of the light curves that show white noise behavior (i.e., $|\text{Kurtosis}| > 10, |\text{Skewness}| > 1, \text{and Std}>0.1$). Removing white noise samples avoids training on a dataset dominated by irrelevant information while keeping variable objects that might contain informative variability. After this filtering process, $\num{1529386}$ light curves are left, with a cadence mean of $2.9\pm 17.3$ days, which fits the lower bound of the PE wavelengths mentioned at the end of Section \ref{sec:pe}.

\subsection{Labeled datasets}\label{sec:labeled_data}
We employ \num{20894} labeled variable stars from the MACHO survey (\citetads{Alcock2001Variable}) to fine-tune and evaluate the downstream task. Table \ref{tab:alcock} shows the class distribution. 

\begin{figure}
    \centering
    \includegraphics[scale=0.75]{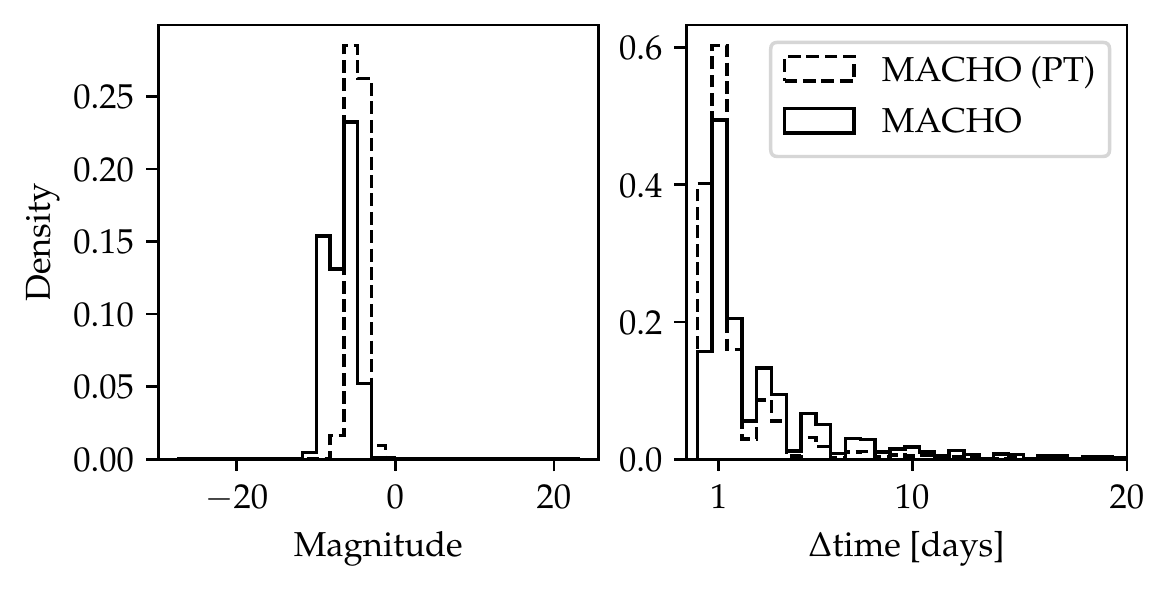}
    \caption{Comparison between labeled and unlabeled pre-training (PT) MACHO dataset. Continuous line represent the labeled MACHO dataset containing variable stars from \citeads{Alcock2001Variable}. [Left] Distribution of light curves magnitudes. [Right] Distribution of the differences between observation times.}
    \label{fig:alcock_vs_macho}
\end{figure}

Like the unlabeled dataset, this catalog contains objects from the Large Magellanic Cloud. Consequently, it contains light curves quite similar to the pre-training ones. However, labeled objects come from all the survey's fields, while the pre-training dataset is a subset (Field 1, 10, 101-104). Figure \ref{fig:alcock_vs_macho} shows the distribution of magnitudes and observational times difference ($\Delta \text{time}$) from the labeled and unlabeled (PT) datasets. In the left histogram, we can see the PT dataset overlaps one of the modes from the labeled dataset distribution. Similarly, distributions of delta times follow the same trend along the $x$-axis. However, they are slightly different for values of delta time greater than one day, where a greater density from the labeled dataset is observed.
The labels were updated to conform to modern classifications. In particular, we group the LPV classes into one category and remove the categories "RR Lyrae and GB blends" and "RRe". The former are excluded because of their small sample size, and the latter because RRe are not considered a sub-class on its own based on later studies (\citet{catelan2004rr}).

We also tested ASTROMER on the Optical Gravitational Lensing Experiment (OGLE-III; \citetads{udalski2004optical}) and ATLAS \citet{heinze2018first} catalogs of variable stars. The OGLE-III data used was selected by \citetads{becker2020scalable} and contains \num{358288} labeled variable stars and correspond to I-band light curves with mean cadence of $3.8 \pm 14.6$ days. Table \ref{tab:ogle} shows its class distribution. The OGLE-III catalog is chosen because its observations are not captured in the same filter as MACHO but close to its wavelength range, as shown in Table \ref{tab:filters}.

\begin{table}
\caption{MACHO catalog distribution.}              
\label{tab:alcock}  
\centering 
\begin{tabular}{c c c} 
\hline\hline         
Tag & Class Name & \# of sources \\ \hline
 Cep\_0 & Cepheid type I &\num{1182} \\
 Cep\_1 &Cepheid type II & \num{683} \\
 EC &Eclipsing binary & \num{6824} \\
 LPV &Long period variable &  \num{3046} \\
 RRab &RR Lyrae type ab  &  \num{7397} \\
 RRc &RR Lyrae type c &  \num{1762} \\
 Total & & \textbf{\num{20894}} \\\hline
\hline                            
\end{tabular}
\end{table}
\begin{table}
\caption{OGLE-III catalog distribution}    
\label{tab:ogle}
\centering 
\begin{tabular}{c c c} 
\hline\hline         
Tag & Class Name & \# of sources \\
\hline
EC &Eclipsing binary &  \num{6862} \\
ED &Detached Binary &  \num{21503} \\ 
ESD &Semi-detached Binary &  \num{9475} \\
Mira &Mira &  \num{6090} \\
OSARG &Small-amplitude red giant  &  \num{234932} \\
RRab &RRLyra type ab &  \num{25943} \\
 RRc &RRLyra type c  & \num{7990} \\
SRV &Semi-regular variable &  \num{34835} \\
cep & Cepheid & \num{7836} \\
dsct &Delta Scuti &  \num{2822} \\
Total & & \textbf{\num{358288}} \\
\hline                            
\end{tabular}
\end{table}
\begin{table}
\caption{Surveys and the filters used in this work.}
\label{tab:filters}    
\centering                                      
\begin{tabular}{c c c c}          
\hline\hline                        
Survey & Filter Name & $\lambda_{min}$\; $\circ{A}$ & $\lambda_{max}\; \circ{A}$ \\  
\hline                                   
MACHO$^{(1)}$ & R &6300 &7600 \\
OGLE$^{(2)}$  & I & 7270 & 8750\\
ATLAS$^{(3)}$ & orange & 5620 & 8200 \\ \hline                                             
\end{tabular}
\tablefoot{In \citet{szymanski2011optical} is noted that OGLE-III and OGLE-IV don't have the same I filters, exhibiting a wide transparency “wing” towards the infrared part of the spectrum. Although different, we use \citet{udalski2015ogle} numbers as a reference, as they are intended to show the approximate ranges of the filters.}
\tablebib{
(1)\citet{alcock1999calibration}, (2)\citet{udalski2015ogle}, (3)\citet{tonry2018atlas}
}
\end{table}
\begin{table}
\caption{ATLAS catalog distribution}              
\label{tab:ATLAS}
\centering 
\begin{tabular}{c c c} 
\hline\hline         
Tag & Class Name & \# of sources \\
\hline
CB & Close Binaries &  \num{80218} \\
DB & Detached Binary &  \num{28767} \\
Mira & Mira &  \num{7370} \\
Pulse &RR Lyrae, $\delta$-Scuti, Cepheids &  \num{25021} \\
Total & & \textbf{\num{141376}}\\
\hline                            
\end{tabular}
\end{table}

The ATLAS dataset, published by \citet{heinze2018first} contains labeled and unclassified objects as well as a dubious class which amounts to roughly $10\%$ actual variable stars and $90\%$ instrumental noise, according to their estimates. From this dataset, \num{422630} light curves are used, measured in the orange passband, as seen in Table \ref{tab:ATLAS}, with a mean cadence of $4.7 \pm 19.1$ days.
The reported classes are grouped to obtain labels similar to the other datasets. In particular, we group CBF and CBH into Close binaries and DBF and DBH into Detached binaries. We do not use the remaining objects, as their labels are based on Fourier analysis and do not correspond exactly to astrophysical categories.

%
%
\subsection{Preprocessing}\label{sec:preprocessing}
\begin{figure}
    \centering
    \includegraphics[scale=1]{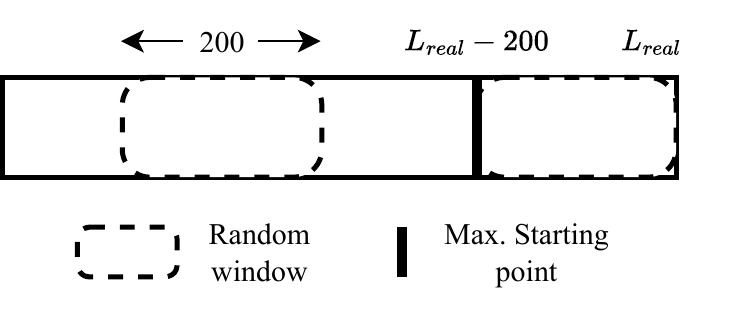}
    \caption{Windows sampling diagram. The rectangle represents the entire light curve whose length is $L_{real}$. The dashed line denotes the sampled windows of size 200 observations. Windows are randomly generated along the light curve. However, We constrain the starting point to be smaller than $L_{real}-200$.}
    \label{fig:windows_gen}
\end{figure}

Neural networks must work with tensors holding equal-length samples. It is not the case of light curves that differ on the number of measurements. As such, it is necessary to set a maximum number of observations for all sequences, padding with zero values if necessary. 
Since we do not want to constrain the encoder to learn from long series exclusively, and 99.52\%  of the unlabeled dataset is longer than 200 (see Appendix \ref{ap:winsize}), we set 200 as the maximum light curve length to be fed to the model. If the light curve is longer than 200, we sample temporal windows starting from a random position, at least 200 observations behind the last measurement. Figure \ref{fig:windows_gen} shows a diagram that exemplifies the process.
On the other hand, if the light curve has less than 200 observations, we pad it with zero values after the last point in order to obtain a sequence of length 200. Those filler observations are masked during training to be excluded from both the attention vector and the loss function.

After generating windows, we independently subtract the mean from each sample. It implies both magnitude and time vectors have zero-mean. We do not scale windows by their standard deviation, either magnitudes or time. Scaling by time dispersion would lose some of the interpretability for the positional encoder. Standardizing magnitudes may lose amplitude-related information, important to discriminate between some classes.

\section{Training Strategy}\label{sec:training_strategy}
The following sections describe the training strategy consisting of two steps: pre-training and fine-tuning. In the pre-training, 60\% of the unlabeled samples are used for training, 20\% for validation, and 20\% for testing. Alternatively, we set 100 samples per class of the labeled dataset to create the test set for evaluating the fine-tuning step. The rest of the labeled dataset is divided into 80\% and 20\% for training and validation, respectively. Recall that the pre-training dataset is different from the labeled ones, so the testing samples used to evaluate the finetuning are never seen by ASTROMER.

Additionally, we study the scientific case of having small target datasets to fine-tune and perform downstream tasks. Therefore, we do not change our 100 samples per class test set but only the training and validation subsets. First, we sample 20, 50, 100, and 500 objects per class from each labeled dataset. Then we divide them by 80/20 for training and validation. We aim to see the impact of the number of samples when adjusting pre-trained weights on a different survey.

\subsection{Pre-training}\label{sec:pretraining}
We implemented ASTROMER using Tensorflow 2 (\citeads{tensorflow2015-whitepaper}). Both pre-training and the other experiments can be reproduced by following the official implementation on Github\footnote{https://github.com/astromer-science}.

The pre-training constitute the first stage of representation learning. It defines the preliminary data requirements, such as cadence and filter, that condition other target domains. In this case, we use the massive unlabeled dataset from the MACHO survey presented in Section \ref{sec:unlabeled}.  

ASTROMER weights are initialized using the Xavier uniform initializer (\citetads{glorot2010understanding}). Training big models such as ASTROMER uses a large amount of hardware resources and computational time. Despite this, once the weights are adjusted, they can be shared, avoiding re-training ASTROMER from scratch.

As mentioned in Section \ref{sec:preprocessing}, the light curves are split into windows of 200 consecutive observations, increasing the effective number of training samples to \num{6201030}.
An epoch is completed after training in all samples once, in batches of 5000. We used Adam (\citetads{kingma2014adam}) with learning rate of $10^{-3}$.

The training performance is evaluated on the validation dataset and the results on the test set. Early stopping is used to stop the training after 40 epochs without improving the validation loss. Once the training is finished, the weights corresponding to the lowest RMSE are saved.

\begin{figure}
    \centering
    \includegraphics[scale=0.75]{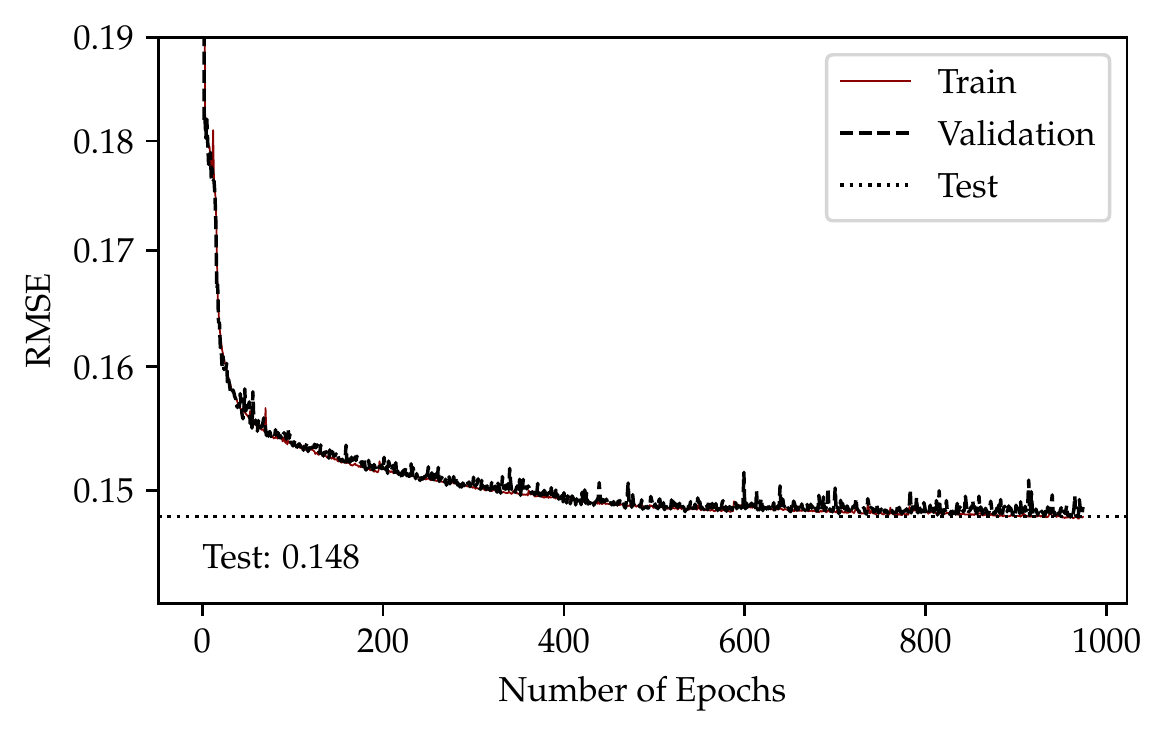}
    \caption{Pre-training learning curves using MACHO unlabeled dataset.}
    \label{fig:pt_results}
\end{figure}

Figure \ref{fig:pt_results} shows the learning curves of the training process associated with the architecture in Figure \ref{fig:arch}. We also trained on smaller models of 64 and 128 attention sizes. However, we chose the 256-dimensional setting as it reached the minimum validation loss on the unlabeled dataset. The model achieved the lowest RMSE close to epoch 1000 \footnote{$\sim 10$ days training on a Nvidia A100 GPU}, obtaining an RMSE of 0.148 on the testing samples (dotted line).

\subsection{Fine-tuning}\label{sec:finetuning}
After pre-training on the MACHO unlabeled dataset, the model has learned much of the variability patterns of the light curves. However, it is still necessary to incorporate downstream task-related information. In this work, we use catalogs of variable stars described in Section \ref{sec:labeled_data} to fine-tune the attention-based embeddings.

The loss function and self-supervised strategy remain the same (i.e., RMSE and masked self-attention technique). Similarly, the same hyper-parameters from the pre-training stage are used.

Table \ref{tab:ft_times} shows the number of epochs and total time to fine-tune ASTROMER on each labeled dataset using all samples, comparing it to the pre-training stage.
A significant difference to the pre-training time is evident. The fine-tuned models converge faster depending on their similarities and the number of examples.
For instance, the model fine-tuned on ATLAS light curves takes more time than the one trained on the MACHO labeled dataset. It is principally due to the survey differences and the number of samples that allow the model to get much lower RMSE than the MACHO labeled dataset. At the same time, similar behavior is observed on the OGLE fine-tuning. However, the minor improvements in OGLE's RMSE suggest the training time is highly dominated by the significant number of samples in the catalog (see Table \ref{tab:ogle}).

\begin{table}
\caption{Fine-tuning results.}.
\label{tab:ft_times}
\centering 
\begin{tabular}{c c c c c} 
\hline\hline         
Dataset & \# Epochs & \# Time &  RMSE\\
\hline
MACHO (PT) & 970 & 9 days 17 hrs & -/0.15 \\\hline
MACHO & 115 & 23 min & 0.09/0.10\\
ATLAS & 147 & 9 hrs 58 min & 0.07/0.22\\
OGLE-III & 244 & 1 day 8 hrs & 0.06/0.08  \\
\hline       
\end{tabular}
\tablefoot{As a reference, the first row shows the performance of the pre-trained model used to initialize weights in the fine-tuning. From the second row forward, the first column indicates the name of the labeled dataset used to fine-tune ASTROMER. The second and third columns show the number of epochs and training time the models spend to converge. The last column is the testing RMSE evaluated on the fine-tuned (left) and pre-trained model (right). For the fine-tuning metrics, we employ balanced testing with 100 objects per class from each labeled dataset}
\end{table}

\begin{figure}
    \centering
    \includegraphics[scale=0.85]{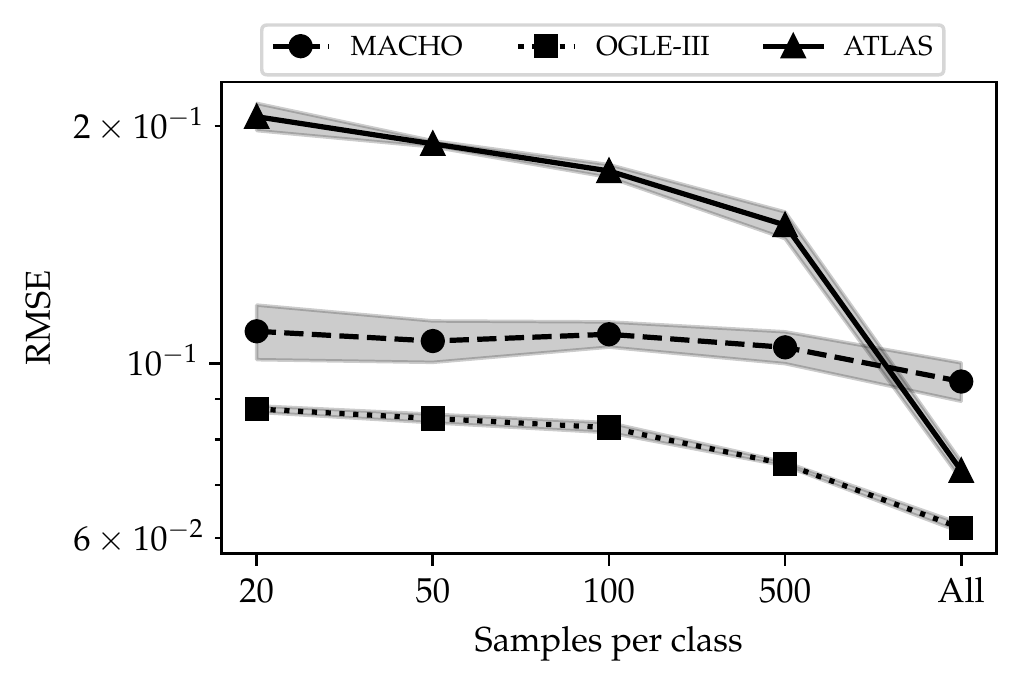}
    \caption{Test RMSE from models fine-tuned on subsets of labeled datasets.}
    \label{fig:rmse_subsets}
\end{figure}

On the other hand, Figure \ref{fig:rmse_subsets} shows the RMSE for the models fine-tuned on smaller datasets of 20, 50, 100, and 500 labels per class. The testing set contains the same 100 objects per class, independent of the training subsets. We can see a descending trend in the RMSE while increasing the number of samples in the fine-tuning, which is expected. However, there is not much improvement along the training subsets in the MACHO labeled dataset since it comes from the same survey as the pre-training light curves. Validation learning curves associated with the training process in all datasets can be found in Appendix \ref{ap:val_lc}.
\section{Specific Task: Classification}
\begin{figure}
    \centering
    \includegraphics[scale=0.84]{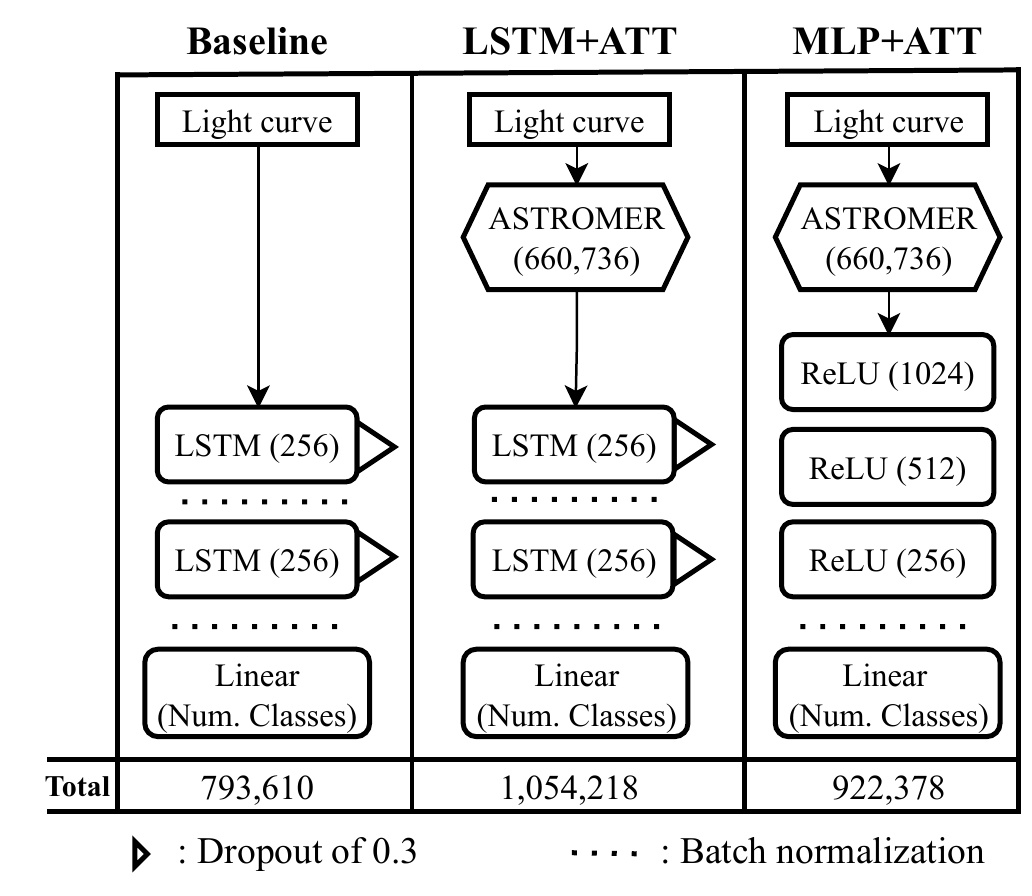}
    \caption{Classifiers architectures. Architecture names with the +ATT tag refer to models trained on ASTROMER embeddings. In contrast, the Baseline uses only the light curves for training. The number in parenthesis is the number of units from the corresponding layer. Notice that the 256 in the LSTM corresponds to the states' size of the RNN. ReLU and Linear corresponds to the activations for each feed-forward layer. The dimensionality of the last linear layer depends on the number of classes from each labeled dataset. The total number of parameter at the bottom does not include the ASTROMER size.}
    \label{fig:classifiers}
\end{figure}

ASTROMER is designed as a general embedding extractor, that can be fine-tuned to solve downstream tasks such as classification or regression. In this work, we cover the problem of classifying variable stars from different surveys. As explained in Section \ref{sec:training_strategy}, the model is fine-tuned in subsets of 20, 50, 100, and 500 samples per class, as well as using the full catalog (See Section \ref{sec:data}). Fine-tuned models are then used to transform light curves into attention-based embeddings. In practice, we use ASTROMER at the beginning of the classifiers as an additional layer (see Figure \ref{fig:classifiers}).

We selected a baseline model from the literature (\citetads{donoso2021effect}) to compare the benefits of using the embeddings as opposed to other deep learning approaches. It consists of two layers of Long Short Term Memory (LSTM) with a memory cell and hidden state of 256 units each. Following the original architecture, we normalize each recurrent layer's output \citepads{ba2016layer} and then apply a dropout \citepads{semeniuta2016recurrent} of 0.3 over them. We do not validate this architecture since it was already tested in the previous work. However, we looked for the best learning rate on MACHO light curves since we changed conditions, such as the batch size and normalization techniques. As shown in Appendix \ref{ap:lr_lstm}, we confirm that a $10^{-3}$ learning rate achieved the best performance in terms of the validation error. In the results shown in Figure \ref{fig:classifiers}, we employed the same LSTM architecture but training with ASTROMER embeddings.

In addition to the LSTM classifier, we tested a Multilayer Perceptron (MLP+ATT in Figure \ref{fig:classifiers}) with three feed-forward hidden layers of 1024, 512, and 256 neurons, and used a Rectified Linear activation function (ReLU) in each of them. Since MLPs cannot process time, we use the average along the steps dimension ($L=200$) of the embedding matrix\footnote{We also tried taking the first, last, and arbitrary attention vectors to feed the input of MLP. However, the results were not better than using the average.}. By doing this, we collapse all the encoded observations in a single vector of size 256 that fits the input dimensionality of the first hidden layer of the MLP+ATT classifier (Feed-Forward layer with 1024 units from Figure \ref{fig:classifiers}).

For small datasets, freezing the weights of ASTROMER allows to optimize a classifier without training a huge number of weights. The ASTROMER encoder can also be trained in tandem with the classification network, to better tune weights to solve a specific task, which can be seen as another fine-tuning step. The following experiments evaluate these two approaches, i.e., training and non-training of the encoder layer of ASTROMER.

To measure the classification performance, we use the F1 score metric,
\begin{eqnarray}
    \textrm{F1}  = \frac{1}{K} \sum_{k=0}^{K-1} 2 \times \frac{\textrm{Precision}_k \times \textrm{Recall}_k}{\textrm{Precision}_k + \textrm{Recall}_k}
\end{eqnarray}
where K is the number of classes, and 
\begin{eqnarray}\label{eq:rec_pre}
    \text{Recall}_k  =   \frac{\textrm{TP}_k}{\textrm{TP}_k + \textrm{FN}_k}\ 
    \text{Precision}_k  =    \frac{\textrm{TP}_k}{\textrm{TP}_k + \textrm{FP}_k}.
\end{eqnarray}
In Equation \ref{eq:rec_pre}, TP, FN, and FP are true positives, false negatives, and false positives cases, respectively. Intuitively, the precision score identifies how many predicted classes are actually valid, and the recall indicates the number of objects in the testing set the model could identify correctly.

\begin{figure*}
    \centering
    \includegraphics[scale=0.85]{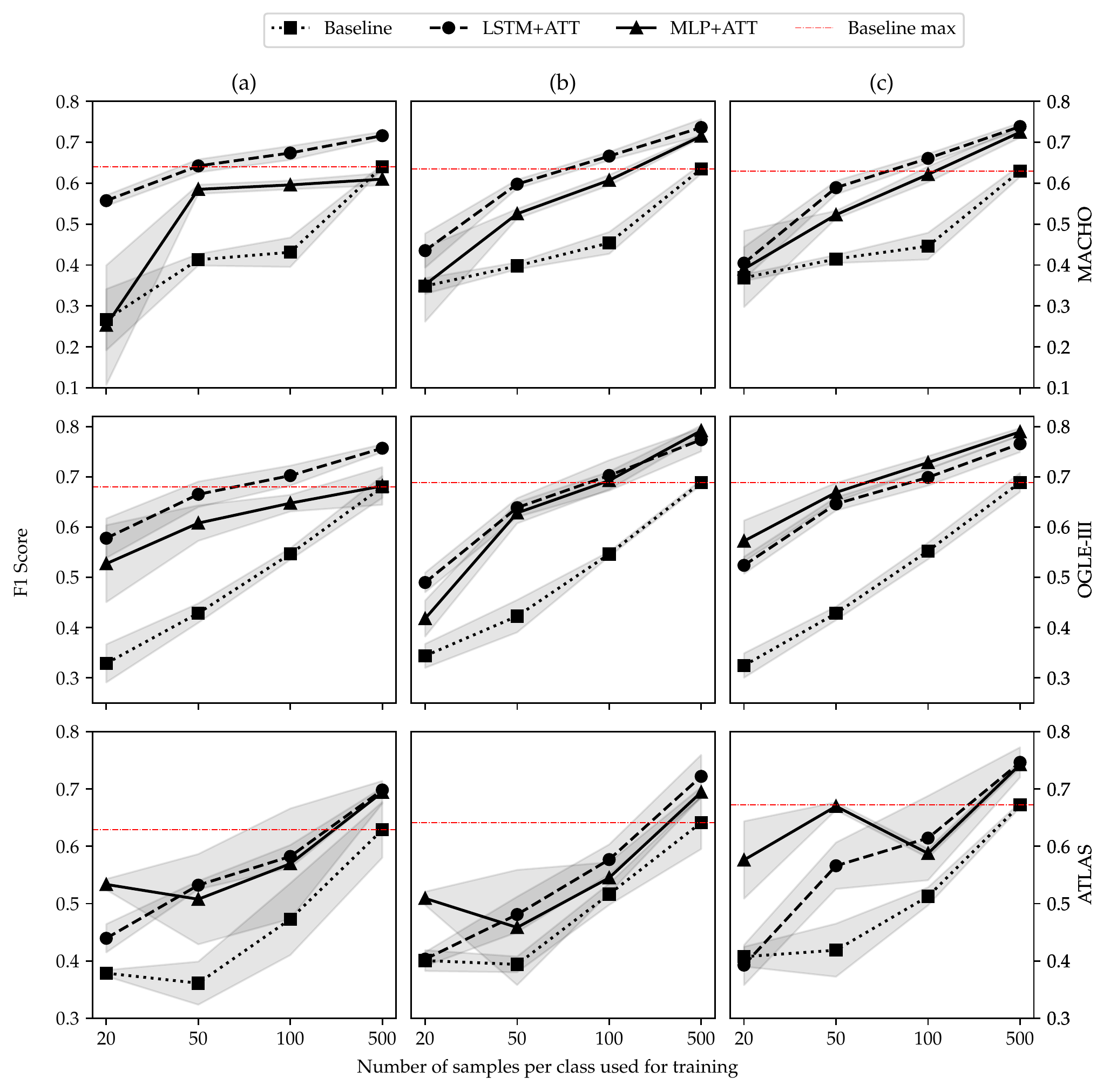}
    \caption{Testing F1 scores for an LSTM trained on light curves directly (Baseline) and models trained on ASTROMER embeddings (LSTM+ATT and MLP+ATT). The gray shading represents the standard deviation of the three cross-validation split. Each row corresponds to the experiments on each survey, MACHO, OGLE and ATLAS, respectively. In (a), we fine-tune ASTROMER and optimize classifiers on smaller subsets of 20, 50, 100, and 500 samples per class. The weights of ASTROMER are kept frozen when classifying. However, in (b), we allow gradients to flow into ASTROMER. The third case (c) shows the results of fine-tuning with the entire set of light curves and classifying on smaller subsets, training ASTROMER simultaneously. 
    }
    \label{fig:subsets_exp}
\end{figure*}

Figure \ref{fig:subsets_exp} shows the scores obtained by the classifiers by (a) freezing the encoder and (b) training the encoder. In addition, a third case (c) is included where all the light curves in the dataset were used to perform fine-tuning, but still doing the classification on the smaller subsets. As in experiment (b), the third column allows gradients to flow into the ASTROMER encoder while training the classifiers. These three scenarios explore often-used strategies when implementing pre-trained models.

The results show that models trained on ASTROMER embeddings perform better than the Baseline in MACHO and OGLE-III across all the experiments. For ATLAS, the difference is smaller but not worse than the Baseline. 
In general, the LSTM trained on attention vectors performs better than the MLP+ATT. However, the MLP+ATT outperforms the Baseline and approaches the LSTM performance as the number of samples per class is increased, in all datasets. 

Note that in (c), even fine-tuning with all the light curves, the score improvement is marginal compared to (b). Similarly, minor improvements can be seen in the classification metrics when optimizing the encoder layer instead of freezing it -i.e., columns (b) and (c) in Figure \ref{fig:subsets_exp}. The MLP+ATT classifier shows the most notorious gain when training with more than 100 samples.

\begin{figure}
    \centering
    \includegraphics[scale=0.8]{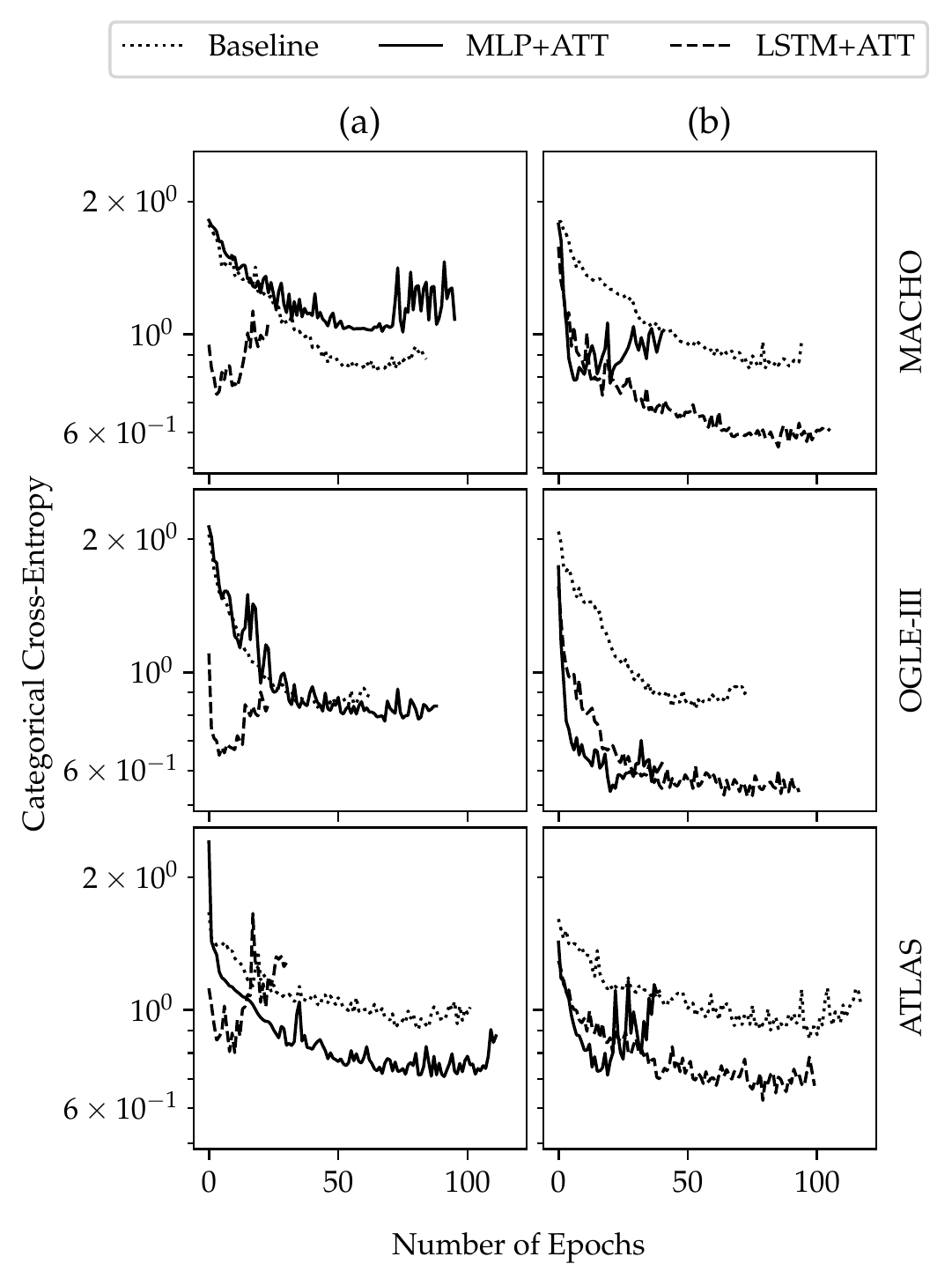}\\
    \caption{Best model validation learning curves for classifiers trained on 500 samples per class on each catalog. The columns show the science cases, i.e., freezing (a) and training (b) ASTROMER when classifying.}
    \label{fig:learning_curves}
\end{figure}

We evaluate the improved learning speed when using ASTROMER. Figure \ref{fig:learning_curves} shows the validation loss associated with the best classifier among the three folds for each dataset and the training scenarios -i.e., freezing (a) and training (b) the ASTROMER encoder. We evaluate learning speed on the most significant subset, which contains 500 objects per class.

Attention-based models generally take fewer epochs to achieve better results than the Baseline. When the encoder stays frozen, the validation curve of the LSTM+ATT is significantly shorter than the other methods, achieving a high F1 score, as shown in Figure \ref{fig:subsets_exp}. In contrast, when training the encoder, the LSTM+ATT takes almost the same number of epochs as the Baseline. However, according to Figure \ref{fig:subsets_exp}, training the encoder using the LSTM+ATT does not improve the F1 score. The findings are reversed when using the MLP+ATT, in this case, training the encoder decreases the number of epochs while improving the F1 score in larger subsets.

\section{Discussion}\label{sec:discussion}
Our results have shown the benefits of using pre-trained models to learn representations of light curves. We successfully adapted and applied an NLP self-supervised technique (\citeads{devlin2018bert}) to the domain of light curves. It offers a solution for learning representations when labels are in limited quantity. Furthermore, it is an alternative to fully unsupervised models that focus more on clustering than making predictions from the data.

Our representation can be extended to light curves coming from different surveys. Using ASTROMER on new data requires a short training process called fine-tuning to adjust the representation and minimize the RMSE. A lower RMSE implies a better representation and, subsequently, a better result in downstream tasks.

The size of the fine-tuning dataset depends on the similarity between the pre-training and fine-tuning magnitudes and cadence distributions. 
For illustration, figure \ref{fig:subsets_exp}c shows no difference in the F1 scores when all the light curves in the dataset were used to fine-tune the encoder. Indeed is more important to bring the classifier more labels than more light curves in the fine-tuning.
As shown in Table \ref{tab:ft_times}, minor improvements in RMSE were obtained when fine-tuning ASTROMER on MACHO and OGLE-III, both more similar to the pre-training dataset than ATLAS. In this scenario, the fine-tuning stage can even be omitted without affecting the results.

Regarding the downstream task, we demonstrated the power of using embeddings to train classifiers, instead of directly using light curves to predict classes. We evaluated the classification performances in three commonly used scenarios using limited labeled data. As seen in Figure \ref{fig:subsets_exp}, we outperform a recurrent neural network classifier trained on light curves (Baseline) in all the experiments. The improvements in the F1 score became more significant when less than 500 labels per class were provided.

ASTROMER can learn discriminative features from the pre-training and fine-tuning steps using only a self-supervised task. As shown in Figure \ref{fig:subsets_exp}, the LSTM+ATT can predict classes with high F1 scores without the need to train the encoder along the classifier. Furthermore, its performance remains the same if the encoder is trained in tandem, independent of the number of samples per class.

While the LSTM+ATT uses its hidden state to capture long and short-term dependencies on the attention vectors over time, the MLP+ATT deals with their average losing global context information. However, when training the encoder along with MLP+ATT using more than 100 samples per class, the encoder is able to counteract this effect, as shown in Figure \ref{fig:subsets_exp}.

In both models, the discriminative features learned by the ASTROMER encoder allow the classifiers to converge faster\footnote{Note that shorter training times using shared pre-trained representations imply the use of less powerful hardware and fewer resources, decreasing the time and energy consumption when training deep learning models \citepads{dhar2020carbon}}.

From the best performing scenarios, the LSTM+ATT converges faster than the Baseline when freezing the encoder weights, as it only needs to recognize discriminative patterns in the embeddings. On the other hand, the MLP+ATT only needs to mitigate the effect of averaging the attention vectors. This task is simpler than learning a representation from scratch, which is the case of the Baseline classifier.  
Specifically, the Baseline has to learn how to extract an informative representation and learn discriminative patterns simultaneously. In the training process, any change in the representation will impact the subsequent layers, increasing the number of epochs to converge. This behavior is related to the Internal Covariance Shift (\citeads{ioffe2015batch}).

Finally, even though this work's main contribution is to help train downstream models on small datasets, we achieved competitive results against \citetads{donoso2021effect} using OGLE-III unfolded light curves ($88.1\%$ vs. $88.0\%$ ours). 
For the MACHO catalog, we compare ASTROMER-based classifiers against the best ($78.1\%$ accuracy) single-band non-period informed model from \citetads{jamal2020neural}. We achieved similar results, $78.2\%$ and $76.7\%$ classification accuracy for the LSTM+ATT and MLP+ATT, respectively.It is important to mention that all models from the literature use random smaller partitions while we use a balanced testing set of 100 objects per class, which is more representative.

\section{Python Package}\label{sec:package}
We provide a Python package of ASTROMER, which includes the pre-trained weights obtained in section \ref{sec:pretraining}. More pre-trained models will be uploaded in the future, either from the ASTROMER team or the community. 

The stable version \code{0.x.y} matches the code of this paper. Variables \code{x} and \code{y} refer to the minor changes and patches of the current version 0. Major changes will not be related to this work but must be duly evaluated to guarantee at least the same performance. ASTROMER \code{v0} can be easily installed from the Python Package Index (PyPI) repository as follow:
\begin{verbatim}
pip install ASTROMER
\end{verbatim}
The principal module \code{ASTROMER.models} include the models associated to the different encoders. So far, we only have the \code{SingleBandEncoder} corresponding to the ASTROMER model trained on the MACHO unlabeled dataset. To import and use ASTROMER we can type,
\begin{verbatim}
from ASTROMER.models import SingleBandEncoder
model = SingleBandEncoder()
\end{verbatim}
where \code{model} is an new instance of ASTROMER. To load pre-trained weights, we must use the \code{from\_pretraining()} method from the \code{model} object,
\begin{verbatim}
model = model.from_pretraining("macho")
\end{verbatim}
The name in parenthesis matches the zip file name in the public repository\footnote{https://github.com/astromer-science/weights}.

The simplest way to obtain embeddings is via a collection of NumPy arrays, including light curves information in the form,
\begin{verbatim}
data = [ np.array([[5200, 0.3, 0.2],
                   [5300, 0.5, 0.1],
                   [5400, 0.2, 0.3]]),

         np.array([[4200, 0.3, 0.1],
                   [4300, 0.6, 0.3]]) ]
\end{verbatim}
where the axes (from left to right) of the NumPy array are the times, magnitudes and standard deviation of the magnitudes. Then, to get the embeddings we use the \code{encode} method,
\begin{verbatim}
att_vectors = model.encode(data)
\end{verbatim}
where \code{att\_vectors} is a list containing the embeddings for each sample in \code{data}.

ASTROMER can be easily trained using the \code{fit()} method from the \code{SingleBandEncoder} instance.
\begin{verbatim}
model.fit(training_data, 
          validation_data, 
          epochs=10)
\end{verbatim}
Within the fit method, \code{training\_data} and \code{validation\_data} are TensorFlow datasets. We provide a function in the \code{ASTROMER.preprocessing} module to format datasets,
\begin{verbatim}
from ASTROMER.preprocessing import load_numpy

training_data = load_numpy(data,
                           batch_size=2,
                           msk_frac=.5,
                           rnd_frac=.2,
                           same_frac=.2,
                           max_obs=200)
\end{verbatim}

Notice \code{data} is the same collection of NumPy arrays we used in the previous examples. In the \code{load\_numpy()}:
\begin{itemize}
    \item \code{batch\_size}: Number of samples to process in one forward pass of the training
    \item \code{msk\_frac}: Fraction of the sequence to be masked
    \item \code{rnd\_frac}: Fraction of the mask to be replaced with random values (see Section \ref{sec:building_rep})
    \item \code{rnd\_frac}: Fraction of the mask to be replaced with actual values (see Section \ref{sec:building_rep})
    \item \code{max\_obs}: Maximum windows length (see Section \ref{sec:preprocessing})
\end{itemize}

For more information we open project repositories \footnote{https://github.com/astromer-science} where data, tutorials, documentation, and contributions via issuing pull requests can be found. These contributions can be in the form of functionalities or pre-trained models.
\section{Conclusion} \label{sec:conclusion}
We present ASTROMER, a single-band embedding for light curve representation. It is based on the BERT NLP model, which codifies observations into attention vectors via self-supervised learning, taking advantage of the massive unlabeled volume of data. In this work, we pretrain ASTROMER on millions of R-band light curves from the MACHO survey.

ASTROMER can be fine-tuned on specific domain datasets to solve downstream tasks, such as classification or regression. Here, we use labeled catalogs from MACHO, OGLE-III, and ATLAS surveys to evaluate the effect of using embeddings on the classification of variable stars. By training a MLP and LSTM classifiers on embeddings, we outperform a baseline LSTM network trained on light curves. We evaluated classification performances of fine-tuned models using 20, 50, 100, and 500 samples per class, obtaining better F1 scores in all experiments. Additionally, we showed that the embeddings-based classifiers achieved competitive scores against state-of-the-art solutions. In terms of training times, models trained on ASTROMER representations took less epochs than the baseline LSTM classifier to reach the lowest validation loss.

Our self-supervised approach only considers the reconstruction of magnitudes, not including other tasks, such as the next sentence prediction applied in BERT. As a result, the final representation may not satisfy other downstream tasks correctly. For instance, the consequences of losing global context information were evidenced on the difference between the MLP and LSTM classifier, both using attention vectors. In order to make embeddings more informative, we will explore adding new tasks during the pre-training and fine-tuning steps in future work.

Finally, we provided a python library including MACHO pre-training weights and fine-tuned models used in this work. Moreover, our library allows users to pre-train, fine-tune, and get embeddings on new single-band datasets. We aim to create a collaborative research environment where pre-trained ASTROMER weights can be shared, saving computational resources and improving state-of-the-art models.

\begin{acknowledgements}
This research was supported by the ANID Millennium Science Initiative ICN12 009, awarded to the Millennium Institute of Astrophysics;  FONDECYT Initiation Nº 11191130 (G.C.V.); the Patagón supercomputer of Universidad Austral de Chile (FONDEQUIP EQM180042) and CONICYT-PFCHA/Doctorado Nacional/2018-21181990.
\end{acknowledgements}

\bibliographystyle{aa} 
\bibliography{main}

\begin{appendix}
\section{Summing magnitudes to PE}
\begin{figure}[!h]
    \label{ap:input_pe}
    \centering
    \includegraphics[scale=0.8]{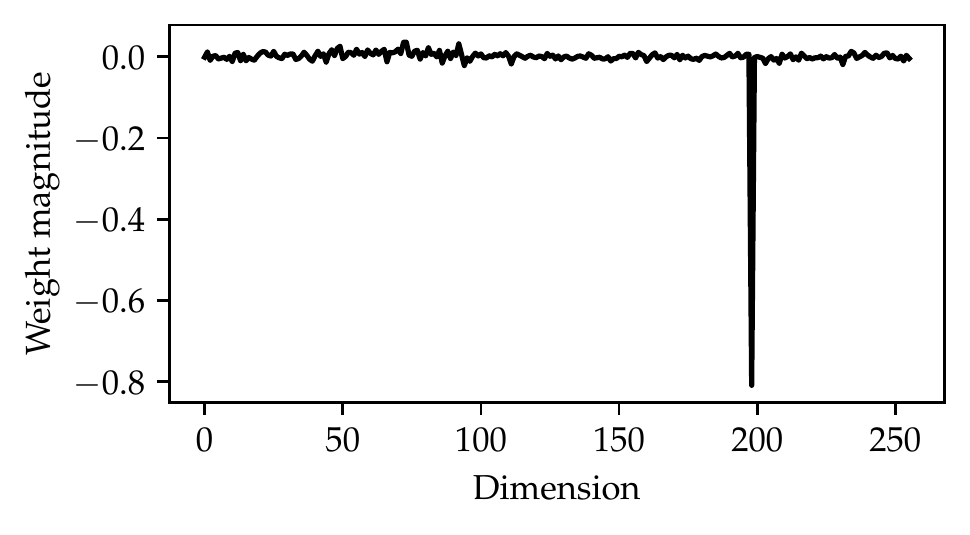}\\
    \caption{Feed-forward network weights that project magnitude scalars to 256-dimensional vectors. The new magnitude vector is then summed to the PE information of equal size.}
\end{figure}

According to the classical approach (\citeads{vaswani2017attention}, \citeads{devlin2018bert}), magnitudes and times should be added to build the input of self-attention blocks. However, summing magnitudes to the PE information is not straightforward as we can interfere with the encoded variability of times. In principle, most of the values in the last dimensions of the PE are constant so that we can sum magnitudes on the unused part (see Figure \ref{fig:pe}). Instead of imposing the above assumption, we let the model learn the way to sum magnitudes from the data. We use a single feed-forward network, transforming magnitude scalars to 256-dim vectors that match the PE dimensionality. After training, we confirmed that most of the projected dimensions are zero, except for one close to 200-th dimension (see Figure \ref{ap:input_pe}). Note that the resulting transformation is almost a one-hot encoding. Thus, we showed that the magnitudes do not interfere with the PE information when joining both vectors.

\section{Looking for a window size}\label{ap:winsize}
Our pre-processing pipeline consists of windows that move along the light curve, sampling a subset of observations. Using this technique, we deal with the variable length problem while generating more samples for training. Figure \ref{ap:winsize} shows the distribution of the lengths of the light curves. Considering that most samples are longer than 200 observations, we defined 200 as the window size.
\begin{figure}[!h]
    \centering
    \includegraphics[scale=0.7]{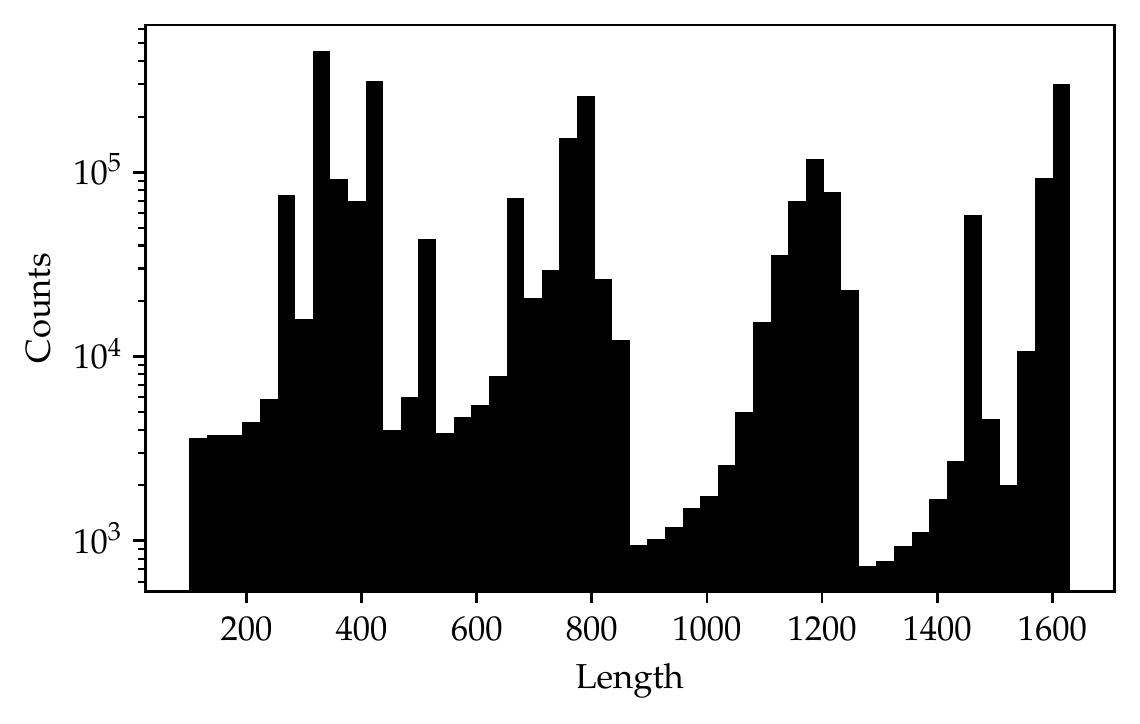}\\
    \caption{Distributions of light curves lengths from the MACHO pre-training dataset.}
\end{figure}
 
\section{Finetuning validation light curves}
\begin{figure*}\label{ap:val_lc}
    \centering
    \includegraphics[scale=0.6]{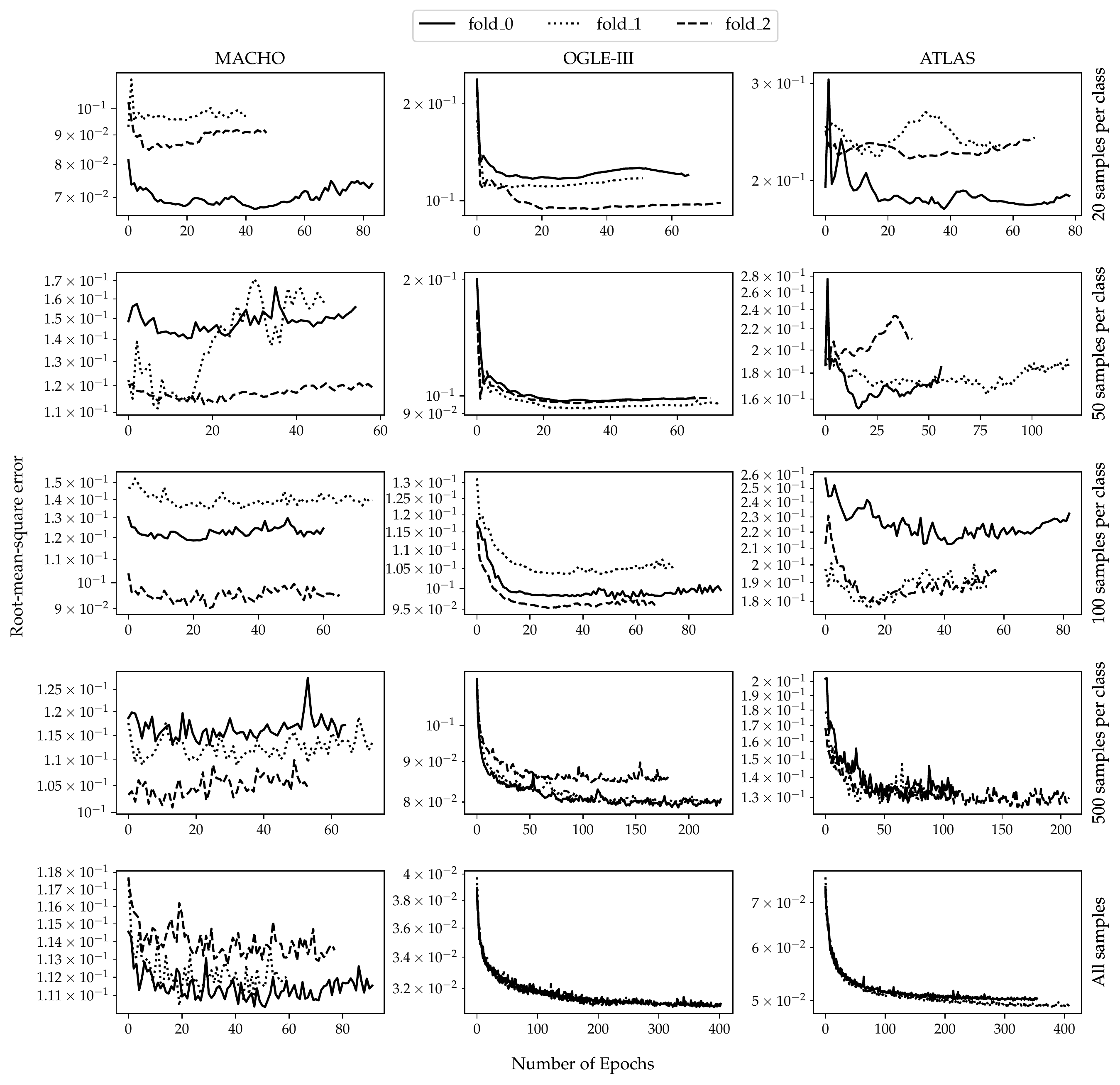}\\
    \caption{Validation learning curves from all datasets during fine-tuning.}
\end{figure*}

Figure \ref{ap:val_lc} shows the validation learning curves associated with ASTROMER fine-tuned on different training sets. The figure rows are associated with subsets of 20, 50, 100, and 500 samples per class. We also show the learning curves for the fine-tuning using all light curves in the respective catalogs in the last row. The different lines within each subplot show the RMSE of the cross-validation splits.
  
\section{Looking for LSTM optimum learning rate}
\begin{figure}
\label{ap:lr_lstm}
    \centering
    \includegraphics[scale=0.8]{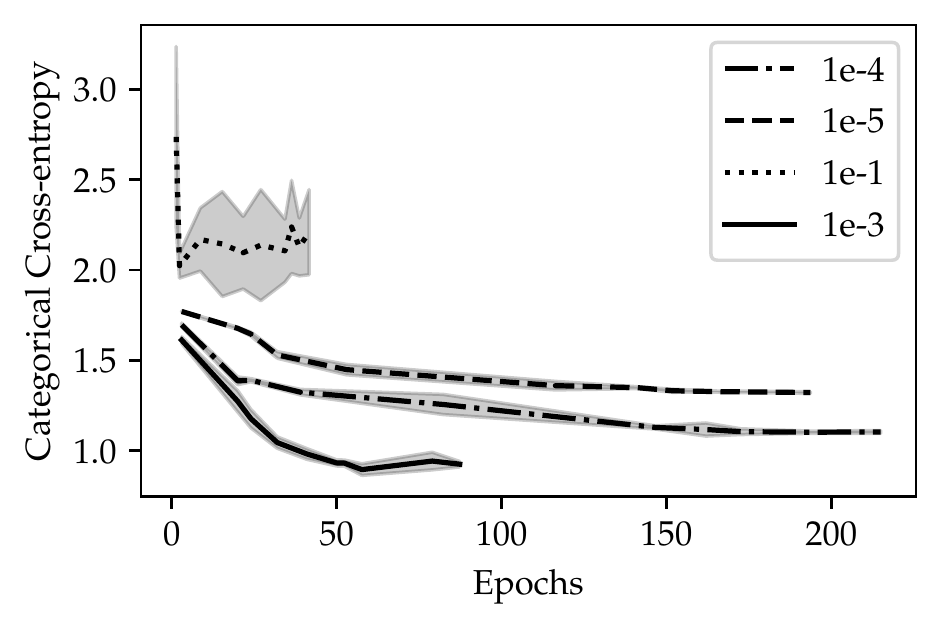}\\
    \caption{Validation learning curves from the Baseline classifier (LSTM) using different learning rates.}
\end{figure}

In order to evaluate the effectiveness of ASTROMER embeddings, we selected a state-of-the-art light curve classifier from the literature \citepads{donoso2021effect}. The classifier consists of two layers of long short-term memory (LSTM) with 256 units each. The idea is to compare how the model performs when changing the input from light curves to light curve embeddings. We kept hyper-parameters as they were defined in the previous work. However, since we changed the batch size and normalization technique, we looked for the best learning rate on MACHO light curves. Figure \ref{ap:lr_lstm} summarizes the cross-validated results showing that a rate of 0.001 performs better in terms of minimal validation loss.

\end{appendix}

\end{document}